\documentclass[aps,prl,twocolumn,showpacs,preprintnumbers,amsmath,amssymb,superscriptaddress]{revtex4}%
\usepackage{graphicx}% Include figure files
\usepackage{dcolumn}% Align table columns on decimal point
\usepackage{bm}% bold math
\usepackage{color}
\begin{document}

\preprint{preprint(\today)}

\title{Oxygen Isotope Effects on Lattice Properties of La$_{2-x}$Ba$_{x}$CuO$_{4}$ ($x$ = 1/8)}

\author{Z.~Guguchia}
\email{zurab.guguchia@psi.ch} 
\affiliation{Physik-Institut der Universit\"{a}t Z\"{u}rich, Winterthurerstrasse 190, CH-8057 Z\"{u}rich, Switzerland}
\affiliation{Laboratory for Muon Spin Spectroscopy, Paul Scherrer Institut, CH-5232
Villigen PSI, Switzerland}

\author{D.~Sheptyakov}
\affiliation{Laboratory for Neutron Scattering and Imaging, Paul Scherrer Institut, CH-5232 Villigen PSI, Switzerland}

%\author{M.~Bendele}
%\affiliation{Physik-Institut der Universit\"{a}t Z\"{u}rich, Winterthurerstrasse 190, CH-8057
%Z\"{u}rich, Switzerland}

\author{E.~Pomjakushina}
\affiliation{Laboratory for Developments and Methods, Paul Scherrer Institut, CH-5232 Villigen PSI, Switzerland}

\author{K.~Conder}
\affiliation{Laboratory for Developments and Methods, Paul Scherrer Institut, CH-5232 Villigen PSI, Switzerland}

%\author{E.~Morenzoni}
%\affiliation{Laboratory for Muon Spin Spectroscopy, Paul Scherrer Institute, CH-5232
%Villigen PSI, Switzerland}

\author{R.~Khasanov}
\affiliation{Laboratory for Muon Spin Spectroscopy, Paul Scherrer Institut, CH-5232
Villigen PSI, Switzerland}

\author{A.~Shengelaya}
\affiliation{Department of Physics, Tbilisi State University,
Chavchavadze 3, GE-0128 Tbilisi, Georgia}

\author{A.~Simon}
\affiliation{Max Planck Institute for Solid State Research, Heisenbergstr.~1, D-70569 Stuttgart, Germany}

\author{A.~Bussmann-Holder}
\affiliation{Max Planck Institute for Solid State Research, Heisenbergstr.~1, D-70569 Stuttgart, Germany}

\author{H.~Keller}
\affiliation{Physik-Institut der Universit\"{a}t Z\"{u}rich,
Winterthurerstrasse 190, CH-8057 Z\"{u}rich, Switzerland}

\begin{abstract}

 A novel negative oxygen-isotope ($^{16}$O/$^{18}$O) effect (OIE) on the low-temperature tetragonal phase transition temperature $T_{\rm LTT}$ was observed 
in  La$_{2-x}$Ba$_{x}$CuO$_{4}$ ($x$ = 1/8) using high-resolution neutron powder diffraction. The corresponding OIE exponent $\alpha_{T_{\rm LTT}}$ = - 0.36(5) has the same sign as $\alpha_{T_{\rm so}}$ = -0.57(6) found for the spin-stripe order temperature $T_{\rm so}$. 
The fact that the LTT transition is accompanied by charge ordering (CO) implies the presence of an OIE also for the CO temperature $T_{\rm co}$. 
Furthermore, a temperature dependent shortening of the $c$-axis with the heavier isotope is observed. These results combined
with model calculations demonstrate that anharmonic electron-lattice interactions are essential for all transitions observed in the stripe phase of cuprates.

\end{abstract}


\maketitle

Since the discovery of high temperature superconductivity in La$_{2-x}$Ba$_{x}$CuO$_{4}$ (LBCO) \cite{Bednorz} numerous intensive studies have revealed a complex interplay between charge, spin, orbital, and lattice degrees of freedom \cite{Jorgensen,Bussmann-Holder}. While the phase diagram of this cuprate family displays a generic complexity, a specific peculiarity is associated with it. In addition to the phase transition from the high temperature tetragonal (HTT) phase with space group $I$4/$mmm$ to the low temperature orthorhombic (LTO) phase  ($Bmab$),  a further transition takes place in LBCO from the LTO to a low temperature tetragonal (LTT) phase ($P$4$_{2}$/$ncm$) which strongly depends on the Ba content $x$ \cite{Axe}. 
The phase transition from HTT to LTO occurs at $T_{\rm LTO}$ ${\simeq}$ 235 K \cite{HuckerPRB} and is accompanied by the tilting of the oxygen CuO$_{\rm 6}$ octahedra around the [110] direction, resulting in buckling distortions of all Cu-O-Cu bonds in the CuO$_{2}$ planes. At $T_{\rm LTT}$ ${\simeq}$ 55 K alternating octahedra tilt along [100] and [010] \cite{HuckerPRB}. This restores the macroscopic fourfold symmetry, but breaks the one of the individual planes where half of the Cu-O-Cu bonds are buckled. 
%As a consequence stripes are rotated by 90$^{o}$ between adjacent planes. 
The structural transition from LTO to LTT is accompanied by a prominent suppression of superconductivity which is largest for $x$ = 1/8 \cite{Moodenbaugh}. This specific doping level is associated with static stripe order and linked to charge-spin-ordering \cite{Crawford2,Luke, Tranquada1,Tranquada2,Abbamonte,Hucker}. 
The static charge order (CO) appears at the same temperature $T_{\rm co}$ ${\simeq}$ $T_{\rm LTT}$  ${\simeq}$ 55 K, and spin order (SO) sets in at $T_{\rm so}$ ${\simeq}$ 35 K \cite{HuckerPRB}.These results provide clear evidence for a subtle interplay between lattice, spin, and charge degrees of freedom. The experimental data indicate that uni-directional stripe-like ordering is common to cuprates \cite{Kivelson,Vojta,Kohsaka}. 
Despite of various attempts \cite{Tranquada2008,Valla,Li,Shen,Berg1,Tranquadareview,Guguchia,GuguchiaPRL}, 
no consensus on the microscopic mechanism of stripe formation and the relevance of stripe correlations for high-temperature superconductivity in cuprates has been achieved.

 The great interest in structural aspects of cuprates is related to the original idea leading to their discovery, namely, 
Jahn-Teller polarons \cite{Bednorz,Jorgensen,Bussmann-Holder}.
Substantial experimental evidences for a strong electron-lattice interaction has been reported for cuprates (see, e.g.,\cite{Mullerisotope,Kellerisotope}), including
the unconventional oxygen isotope ($^{16}$O/$^{18}$O) effects (OIE's) on various quantities \cite{Mullerisotope,Kellerisotope,shengelaya,khasanov2,Lanzara,Rubio,Ronnow}. 
However, remarkable electronic properties of cuprates stimulated alternative proposals in terms of purely electronic pairing mechanisms (see, e.g.~\cite{Berg1}). 
To explore the role of the lattice dynamics on the various ordering temperatures in the stripe phase of cuprates, OIE studies \cite{Mullerisotope,Kellerisotope,shengelaya,khasanov2,Lanzara,Rubio,Ronnow} are essential.

 So far, a large OIE on the superconducting (SC) transition temperature $T_{\rm c}$ was reported for La$_{1.6-x}$Nd$_{0.4}$Sr$_{x}$CuO$_{4}$ \cite{Wang} and  
La$_{1.8-x}$Eu$_{0.2}$Sr$_{x}$CuO$_{4}$ \cite{Sury} showing stripe order at $x$ = 1/8  \cite{Crawford,Buchner}. 
Recently, we observed substantial OIE's on magnetic quantities characterizing the static spin-stripe order in La$_{2-x}$Ba$_{x}$CuO$_{4}$ with $x$ = 1/8 (LBCO-1/8) \cite{GuguchiaPRL}. 
However, no OIE investigations on the LTT phase transition and charge order in 1/8 doped  cuprates have been reported.

\begin{figure}[b!]
\centering
\includegraphics[width=0.9\linewidth]{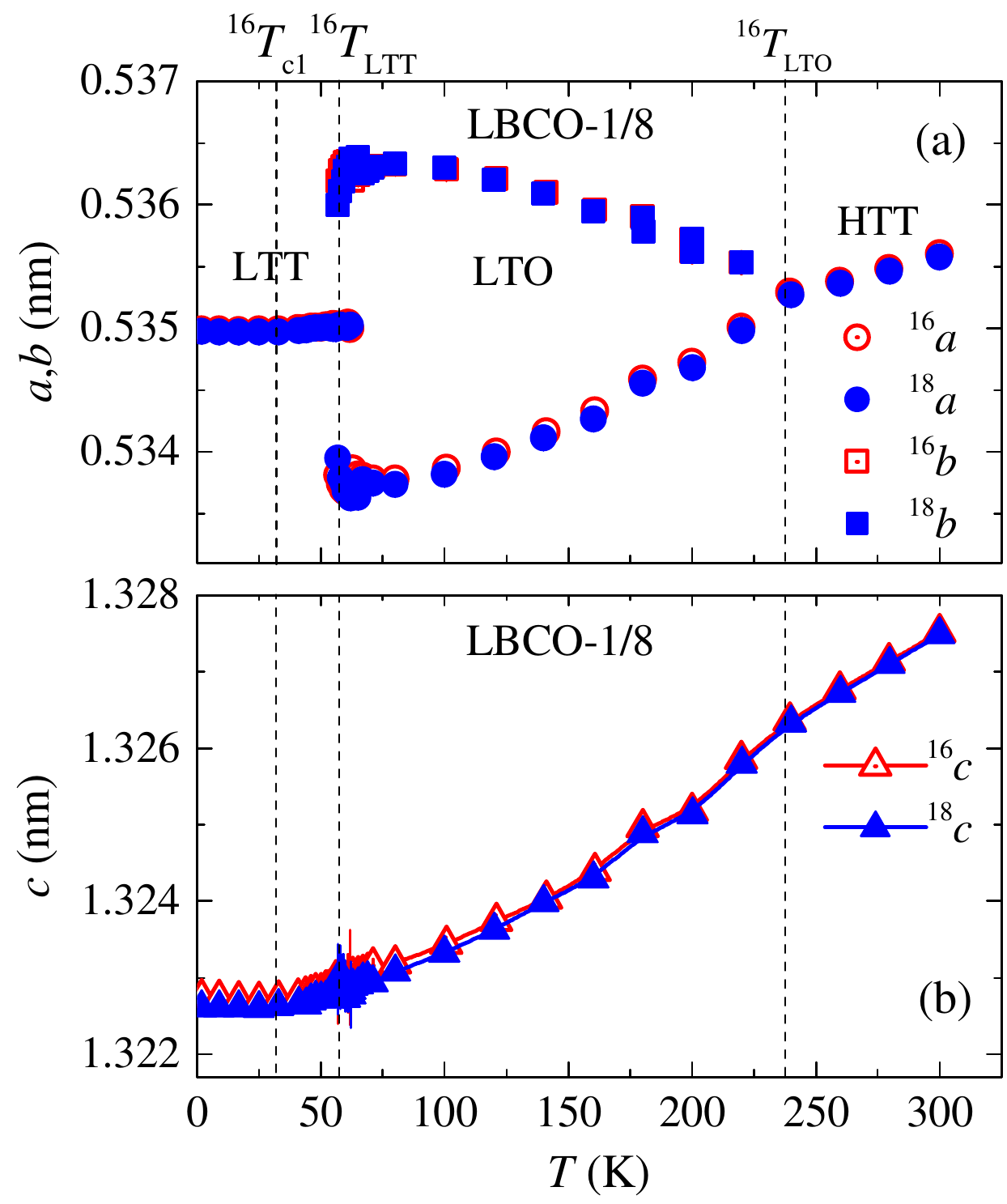}
\vspace{-0.3cm}
\caption{ (Color online) Temperature dependence of the lattice parameters $a$, $b$ (a),  and $c$ (b) for the 
$^{16}$O and $^{18}$O samples of LBCO-1/8. The vertical lines mark the SC transition temperature $^{16}$$T_{\rm c1}$ as well as the structural phase 
transition temperatures $^{16}$$T_{\rm LTT}$ and $^{16}$$T_{\rm LTO}$.}
\label{fig1}
\end{figure}
%%%%%%%%%%%%%%%% 
%%%%%%%%%%%%%%%%%%%%%%%%%%%%%%%%%%%%%%%%%%%%%%%%%%%%%%%%%%%%%%
\begin{figure}[t!]
\centering
\includegraphics[width=0.9\linewidth]{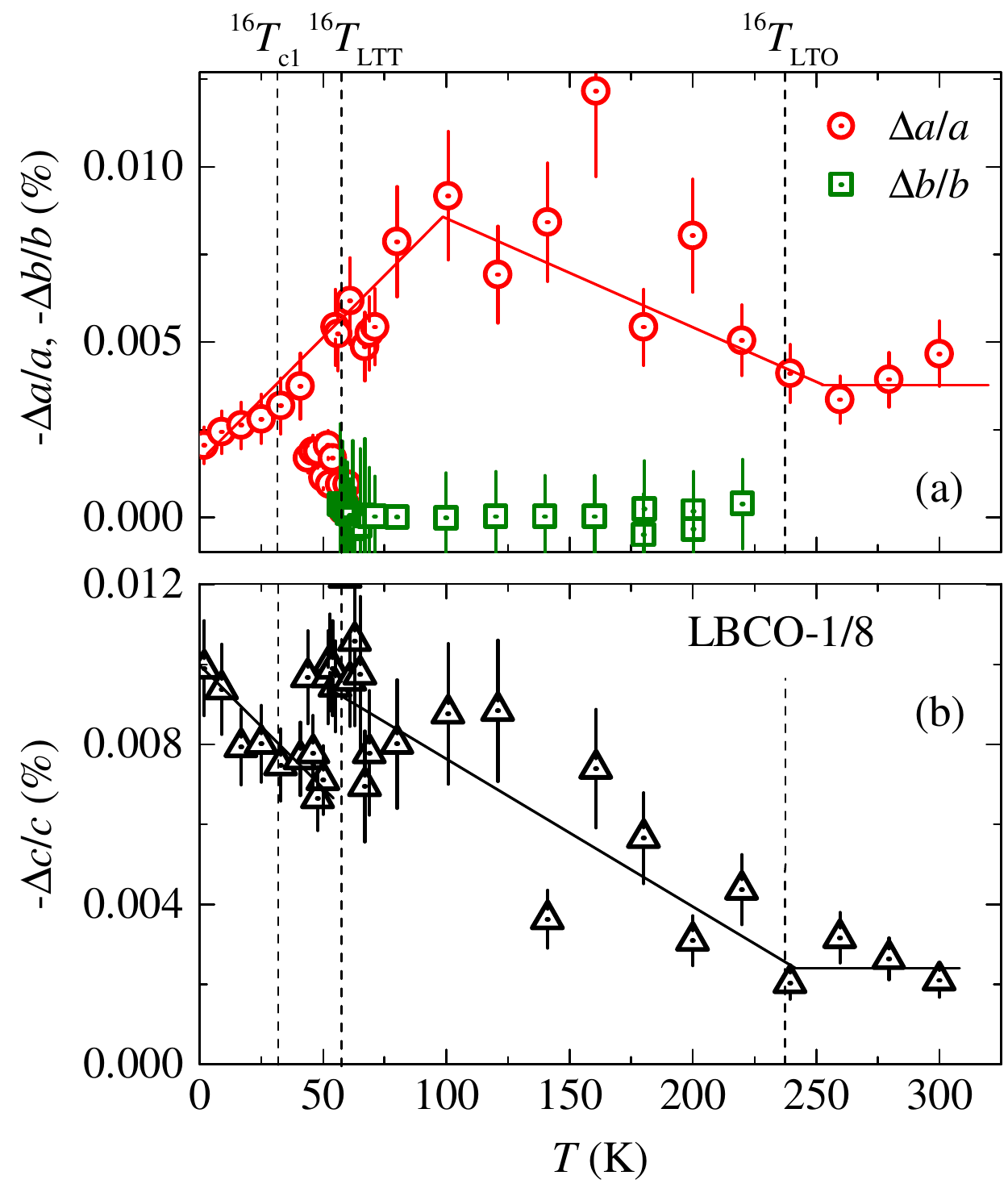}
\vspace{-0.3cm}
\caption{ (Color online) Temperature dependence of the relative oxygen isotope shifts
${\Delta}$$a$/$a$, ${\Delta}$$b$/$b$ (a), and ${\Delta}$$c$/$c$ (c) for the lattice parameters $a$, $b$, and $c$ of LBCO-1/8.  
The vertical lines mark the SC transition temperature $^{16}$$T_{\rm c1}$ as well as the structural phase 
transition temperatures $^{16}$$T_{\rm LTT}$ and $^{16}$$T_{\rm LTO}$.}
\label{fig1}
\end{figure}
%%%%%%%%%%%%%%%% 

 In this letter we report a new negative OIE on $T_{\rm LTT}$ in LBCO-1/8 determined by means of high-resolution neutron powder diffraction with an
OIE exponent $\alpha_{T_{\rm LTT}}$ = - 0.35(5), which has the same sign as $\alpha_{T_{\rm so}}$ = -0.57(6) found for the spin-stripe order temperature $T_{\rm so}$ \cite{GuguchiaPRL}. Since the LTT transition is accompanied by static charge order, we expect an OIE also to be present for the CO temperature. These results establish that all transitions observed in the stripe phase of LBCO-1/8 are sensitive to oxygen ion related lattice vibrations,
indicating that these are relevant for stripe formation in the cuprates.
 
%, demonstrating that the electron-lattice interaction is essential for the stripe formation in the cuprates.

 %  OIE investigations of   
%the static spin-stripe order in LBCO-1/8 by means of ${\mu}$SR experiments. 
%Substantial OIE's were found on magnetic quantities characterizing the static spin-stripe phase,
%demonstrating that the electron-lattice interaction is essential in the stripe formation mechanism of cuprates.
%In addition, we also studied the OIE on $T_{\rm c}$ in  LBCO-1/8 by magnetization measurements.
%Remarkably, it was found that the OIE's have opposite signs for the magnetic and superconducting states 
%in the stripe phase of LBCO-1/8. These findings reveal that lattice vibrations play an important role in the competition  
%between superconductivity and static spin-stripe order in LBCO-1/8.

 New polycrystalline samples of oxygen-substituted ($^{16}$O/$^{18}$O) La$_{2-x}$Ba$_{x}$CuO$_{4}$ with $x$ = 1/8 were prepared in the same way as described in \cite{GuguchiaPRL}. Magnetization, ${\mu}$SR, and powder neutron diffraction experiments were performed on samples from the same batch in order
to study the OIE's on $T_{\rm c1}$ \cite{GuguchiaPRL}, $T_{\rm so}$, and $T_{\rm LTT}$ and to make a direct comparison between the OIE's on various quantities in the stripe phase of LBCO-1/8. 
%Magnetization experiments were performed with  a SQUID magnetometer ($Quantum$ $Design$ MPMS-XL) in a field of 0.5 mT and at temperatures between 2 K and 50 K. 
%The ${\mu}$SR experiments were carried out at the ${\pi}$M3 beam line at the Paul Scherrer Institut (PSI) in Switzerland using 
%the general purpose instrument (GPS) with a standard veto setup providing a low-background ${\mu}$SR signal.
The OIE exponents for $T_{\rm c1}$ and $T_{\rm so}$ are $\alpha_{T_{\rm c}}$ = 0.46(6) and $\alpha_{T_{\rm so}}$ = -0.56(9), respectively (see supplemental materials for details),
being in perfect agreement with our previous results \cite{GuguchiaPRL}.
The neutron diffraction experiments were carried out at the high-resolution
powder diffractometer HRPT \cite{Fischer} at the Swiss spallation neutron source SINQ at the Paul Scherrer Institut (PSI), Switzerland, with a wavelength of ${\lambda}$ = 0.1494 nm. 
\begin{figure}[b!]
\centering
\includegraphics[width=0.9\linewidth]{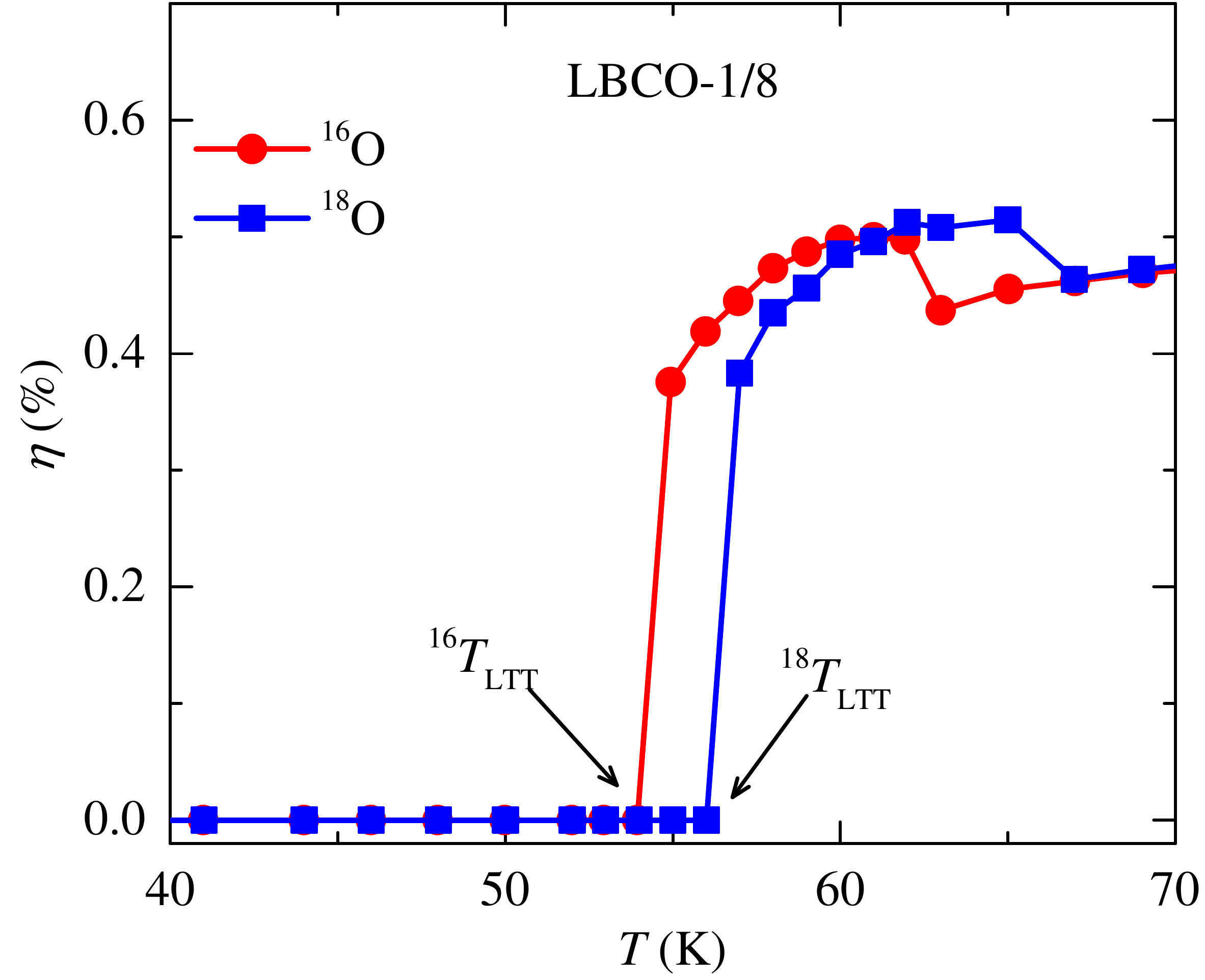}
\vspace{-0.4cm}
\caption{ (Color online) Temperature dependence of the orthorhombic strain ${\eta}$ = 2($a-b$)/($a+b$) for the $^{16}$O and $^{18}$O samples of LBCO-1/8.  The arrows denote the LTT transition temperatures $^{16}$$T_{\rm LTT}$ and $^{18}$$T_{\rm LTT}$ for the $^{16}$O and $^{18}$O samples, respectively. The peculiarities at higher temperatures are artefacts connected with the exchange of the data treatment procedure.}
\label{fig1}
\end{figure}
%%%%%%%%%%%%%%%% 

The OIE's on the LTT structural phase transition temperature $T_{\rm LTT}$ and on the lattice parameters $a$, $b$, and $c$ of LBCO-1/8 were studied by neutron diffraction. 
%Neutron powder diffraction patterns recorded at $T$ = 2 K and the corresponding structural refinements
%for the $^{16}$O  and $^{18}$O samples of LBCO-1/8 are shown in Fig. 2a and b. 
%The average structure was assessed through Rietveld refinements to the raw diffraction data using the program Fullprof \cite{Rodrigez} employing  $I$4/$mmm$ (HTT), $Bmab$ (LTO), and %$P$4$_{2}$/$ncm$ (LTT) models from the literature \cite{HuckerPRB}. 
%Figure 2c and d shows that the orthorhombic splitting between the 400 and 004 Bragg reflections is clearly resolved. Below ${\simeq}$ 55 K the orthorhombic splitting is suppressed, %signaling the transition to the LTT phase.
The crystal structure of LBCO-1/8 at room temperature (RT) was refined with the tetragonal $I$4/$mmm$ space group using the program Fullprof \cite{Rodrigez}. 
Lattice parameters are $a$ = 0.5355(2) nm and  $c$ = 1.3275(7) nm
at RT. The temperature dependence of the lattice parameters for the $^{16}$O  and $^{18}$O 
samples of LBCO-1/8 are shown in Fig.~1a and 1b. The LTO and LTT structural phase transitions are clearly visible in the temperature dependences of $a$ and $b$. 
At $T_{\rm LTO}$ ${\simeq}$ 230 K a splitting of $a$ and $b$ reveals the transition to the LTO structure which smoothly evolves from the HTT phase. Around
$T_{LTT}$ ${\simeq}$ 55 K these lattice parameters merge together, and a first-order transition from the LTO to the LTT phase takes place \cite{Axe}. 
The $c$-axis smoothly contracts with decreasing temperature (see Fig.~1b). 
The values for $T_{\rm LTO}$ and $T_{\rm LTT}$ (marked by the vertical lines in Fig.~1) and those for the lattice parameters 
for the $^{16}$O sample are in good agreement with published values \cite{Axe,HuckerPRB}. 
The OIE's on the lattice constants are displayed in 
Fig.~2a and 2b where the temperature dependence of the relative isotope shifts of the lattice parameters ${\Delta}$$q$/$q$ = ($^{18}$q-$^{16}$q)/$^{16}$q ($q$ = $a$, $b$, $c$)
are shown. In all structural phases the in-plane lattice parameters $a$ and $b$  for LBCO-1/8 decrease only slightly upon replacing $^{16}$O by $^{18}$O. The quantity ${\Delta}$$a$/$a$ increases with decreasing temperature, reaching a maximum around ${\simeq}$ 100 K and decreases below. 
At room temperature ${\Delta}$$c$/$c$ is nearly the same as ${\Delta}$$a$/$a$.  However, upon lowering the temperature ${\Delta}$$c$/$c$ monotonically increases 
to be by a factor of 6 larger at the base temperature than at RT \cite{Comment}. 

%%%%%%%%%%%%%%%%%%%%%%%%%%%%%%%%%%%%%%%%%%%%%%%%%%%%%%%%%%%%%%
\begin{figure}[b!]
\centering
\includegraphics[width=0.9\linewidth]{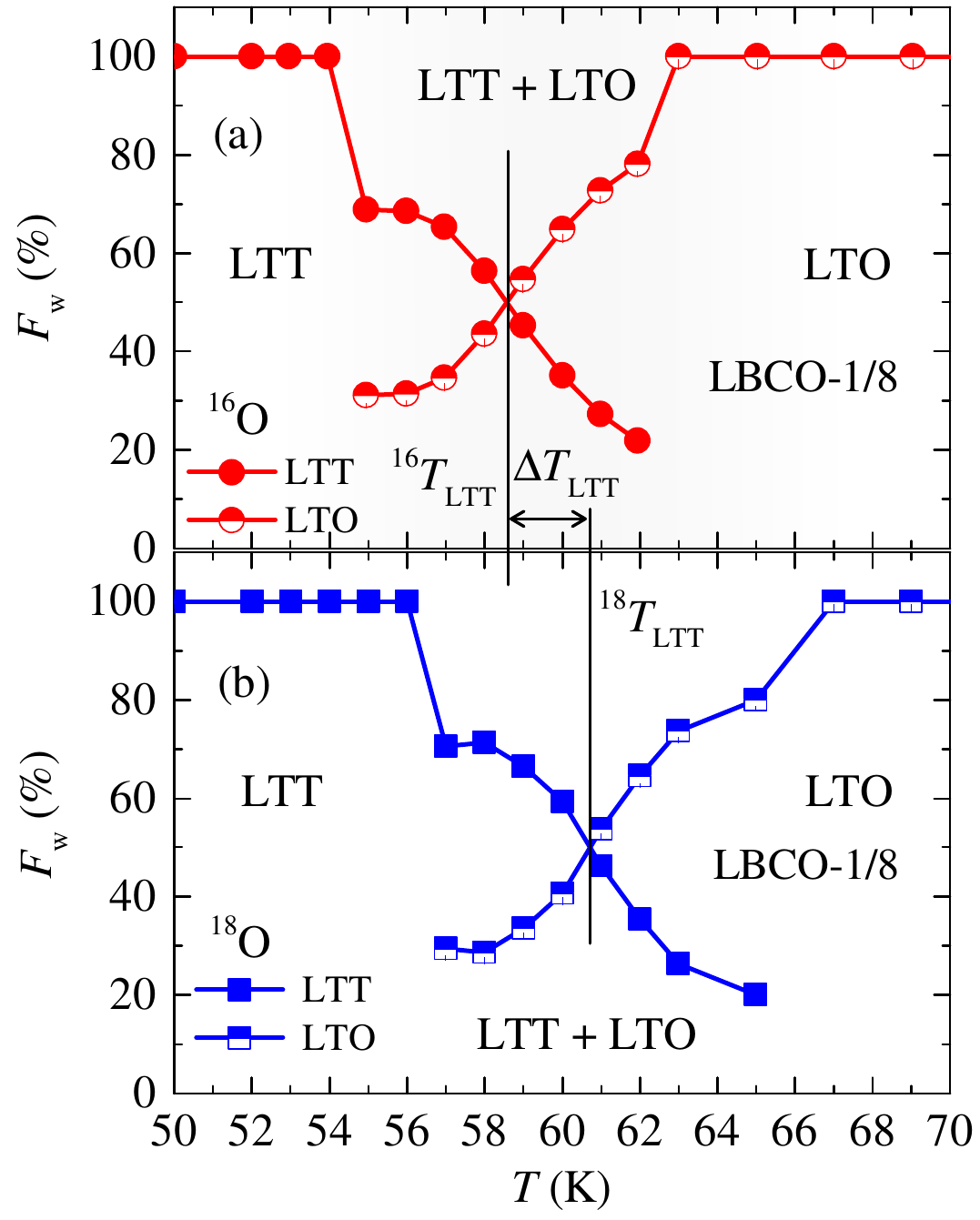}
\vspace{-0.3cm}
\caption{ (Color online) Temperature dependence of the relative weight fraction $F_{\rm w}$ of the LTT and LTO structural phases 
for the $^{16}$O (a) and $^{18}$O (b) samples of LBCO-1/8.
The vertical lines mark the LTT transition temperatures $^{16}$$T_{\rm LTT}$ and $^{18}$$T_{\rm LTT}$ for the $^{16}$O and $^{18}$O samples, respectively.}
\label{fig1}
\end{figure}

In order to illustrate the OIE on $T_{\rm LTT}$, we plot the orthorhombic strain ${\eta}$ = 2(a-b)/(a+b) \cite{HuckerPRB}  for the two isotope samples as a function of temperature (see Fig.~3). ${\eta}$ sharply drops to zero at $T_{LTT}$ ${\simeq}$ 55 K for the $^{16}$O sample, consistent with the first order nature of the LTT transition. The OIE on $T_{\rm LTT}$  is clearly visible as a pronounced shift to higher temperature with increasing isotope mass. The results for the OIE on $T_{\rm LTT}$ are: $^{16}$$T_{\rm LTT}$ = 54.5(3) K, $^{18}$$T_{\rm LTT}$ = 56.5(3) K, ${\Delta}$$T_{\rm LTT}$ = $^{18}$$T_{\rm LTT}$ - $^{16}$$T_{\rm LTT}$ = 2.1(2) K, yielding an OIE exponent $\alpha_{T_{\rm LTT}}$ = -0.36(5).

 %Note that close to LTT transition ${\delta}$($T$) for the 
%$^{18}$O sample is systematically shifted towrads higher temperatures as compared to one for the $^{16}$O sample,
%indicating that  the LTT structural phase transition temperature  $^{18}$$T_{\rm LTT}$ for $^{18}$O
%is higher than $^{16}$$T_{\rm LTT}$ for $^{16}$O. 

%%%%%%%%%%%%%%%%%%%%%%%%%%%%%%%%%%%%%%%%%%%%%%%%%%%%%%%%%%%%%%%
%\begin{figure}[t!]
%\centering
%\includegraphics[width=0.8\linewidth]{Fractionsv4.pdf}
%\vspace{-0.6cm}
%\caption{ (Color online) Temperature dependence of the weight of the LTT and LTO structural phases 
%for the $^{16}$O (a) and $^{18}$O (b) samples of LBCO-1/8.
%The vertical lines mark the LTT transition temperatures $^{16}$$T_{\rm LTT}$ and $^{18}$$T_{\rm LTT}$ for the $^{16}$O and $^{18}$O samples, respectively.}
%\label{fig1}
%\end{figure}
%%%%%%%%%%%%%%%%% 
 The OIE on the LTT transition can also be obtained from the relative weight fractions $F_{\rm w}$ of the LTO and LTT phases as a function of
temperature, evidencing the coexistence of the LTO and the LTT phases over a temperature interval
of ${\simeq}$ 8 K. This is consistent with the first order nature of the LTT transition. 
% Estimates of the relative LTT and LTO fractions are, of course, influenced by assumptions made in the fitting process. We %accurately fit the data   
In Fig.~4a and b we plot the temperature dependences of $F_{\rm w}$ of the LTO and the LTT phase
for the  $^{16}$O and  $^{18}$O samples, respectively. For the  $^{16}$O sample, the LTT phase appears below ${\simeq}$ 63 K and $F_{\rm w}$ increases with decreasing temperature. The transition to the LTT phase is complete below 54 K, and between 54 K and 63 K the LTT phase coexists with the LTO one. 
$T_{\rm LTT}$ is defined as the temperature where $F_{\rm w}$ of the LTO and the LTT phase are equal, yielding 
$^{16}$$T_{\rm LTT}$ = 58.5(3) K, $^{18}$$T_{\rm LTT}$ = 60.7(3) K, and
$\alpha_{T_{\rm LTT}}$ = -0.37(5), in excellent agreement with 
$\alpha_{T_{\rm LTT}}$ = -0.36(5) obtained from the orthorhombic strain (Fig.~3). This demonstrates that the two independent methods for 
the determination of $\alpha_{T_{\rm LTT}}$ yield consistent results. 

The OIE on the structural phase transition temperature $T_{\rm LTT}$ is accompanied by a clear shrinkage of the $c$-axis with the heavier isotope
in the temperature range from 2 K to 300 K (see Fig.~2b). Both phase transitions, the one from HTT to LTO and LTO to LTT are accompanied by substantial phonon softening \cite{Axe,Kimura,Keimer1993}, where the first one is associated with the condensation of a degenerate pair of 
transverse-optic (TO) phonon modes at the X-point wave vector q=(1/2, ±1/2, 0), while the low temperature one is driven by a soft Z-point phonon mode. This mode emerges from the high temperature soft mode, which splits into two modes below $T_{\rm LTO}$, corresponding to a zone boundary mode at the (3,0,2) Z point and a zone center one at the (0,3,2) ${\Gamma}$ point. While the latter of the two modes hardens, the Z point one continues to soften up to $T_{\rm LTT}$. An understanding of this behavior and the transitional properties has been given in terms of a Landau free energy expansion, where symmetry invariant combinations of a degenerate 
order parameter $Q_{\rm i}$ ($i$ = 1,2) and the orthorhombic strain ${\eta}$ have been considered \cite{Axe,Ting}:
%%%%%%%%%%%%%%%%% 
\begin{equation}
\begin{split}
F=1/2a(T-T_0)(Q_1^2+Q_2^2)+u〖(Q_1^2+Q_2^2)〗^2+ \\
v(Q_1^4+Q_2^4 ).... + c{\eta}^2/2 +d(Q_1^2-Q_2^2){\eta}+
\end{split}
\end{equation}
%%%%%%%%%%%%%%%%% 
$T_{\rm 0}$ is the transition temperature which depends on the coefficients $a$, $u$, $v$, $c$, and $d$, and refers either to $T_{\rm HTT}$ or $T_{\rm LTT}$. Within this approach it is possible to construct the complex phase diagram, however, isotope effects are absent. These are obtained within a dynamic anharmonic electron lattice interaction Hamiltonian which reduces in the limit that only the soft modes are taken into account, to \cite{Bussmann-Holder}:
%%%%%%%%%%%%%%%%% 
\begin{equation}
H = \sum_i\frac{p_i^2}{2m}-\frac{g_2}{2 }w_i^2+\frac{g_4}{4}w_i^4+\sum_{i,f} f〖(w_i-w_j)〗^2+...
\end{equation}
%%%%%%%%%%%%%%%%% 
with $p_{\rm i}$ being the momenta in cell $i$, $m$ the oxygen ion mass, $g_{\rm 2}$, $g_{\rm 4}$ the harmonic and fourth order electron-ion interaction constants, $w_{\rm i}$ the relative electron-ion displacement coordinate, and $f$ the long range interactions between sites $i$, $j$. The fourth order term proportional to $g_{\rm 4}$ is treated within the self-consistent phonon approximation (SPA), and a phase transition takes place at \cite{Annette2012}
%%%%%%%%%%%%%%%%% 
\begin{equation}
T_0 = \hbar g_{4}\sqrt{f}[g_{2}^{2}\sqrt{m}~{\rm arccoth} \{2\sqrt{fm} g_{2}/(3g_{4}\hbar)\}]^{-1} 
\end{equation}
%%%%%%%%%%%%%%%%% 
%%%%%%%%%%%%%%%%%%%%%%%%%%%%%%%%%%%%%%%%%%%%%%%%%%%%%%%%%%%%%%
\begin{figure}[t!]
\centering
\includegraphics[width=0.9\linewidth]{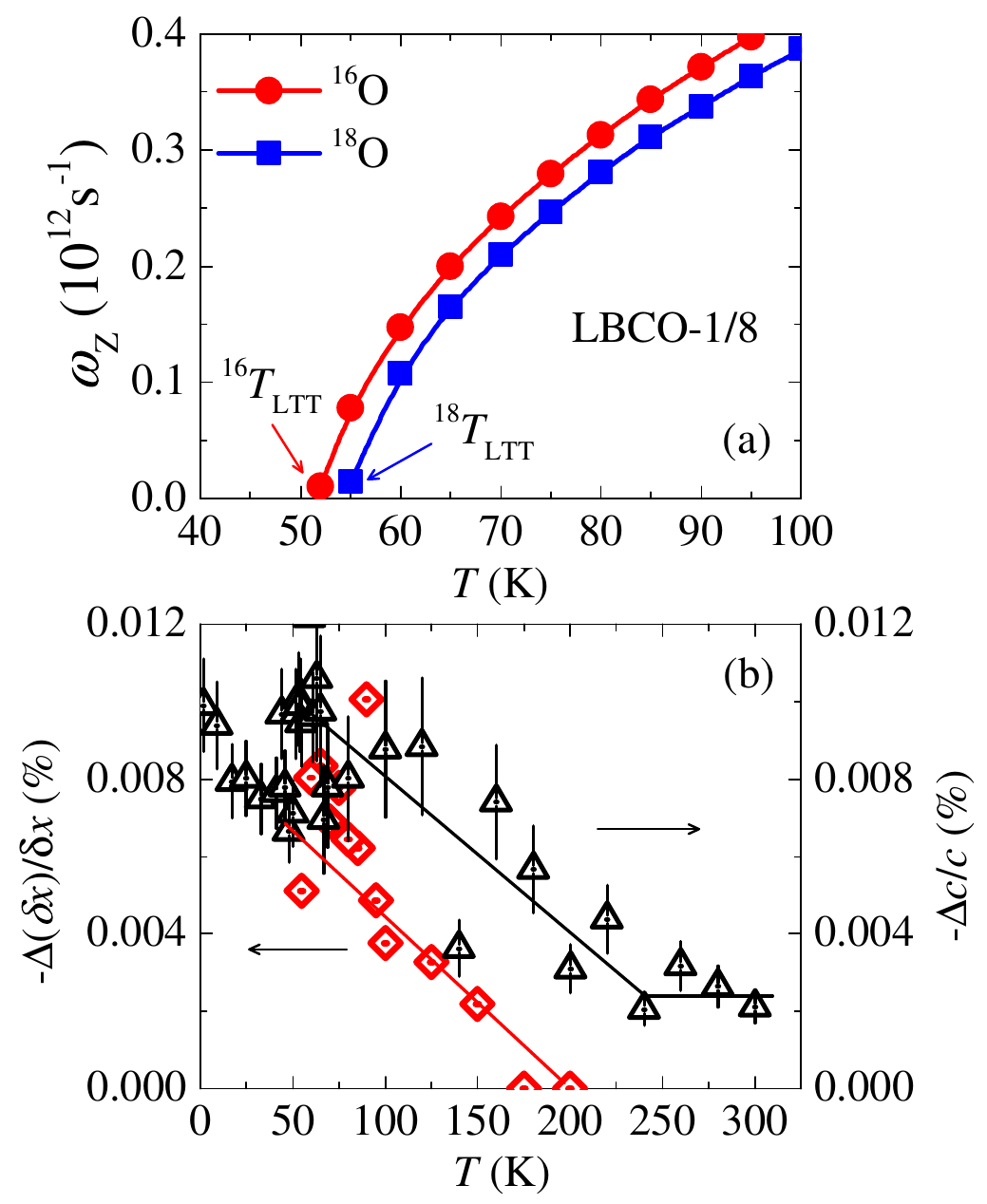}
\vspace{-0.3cm}
\caption{ (Color online) Calculated temperature dependence of the Z-point related phonon mode frequency $\omega_{\rm Z}$ (a)
and the relative isotope shift of the dynamical relative oxygen-copper displacement $\Delta(\delta{x})$/$\delta{x}$ (b) of LBCO-1/8, 
plotted together with ${\Delta}$$c$/$c$, taken from Fig.~2b.}
\label{fig1}
\end{figure}
%%%%%%%%%%%%%%%% 

In order to determine the correct symmetry of the soft mode, the free energy [Eq.~(1)] has to be employed for the corresponding order parameter. This is included in Eq.~(2) 
by solving the equations of motion and take the zero solution for the soft mode at the appropriate $q$-value from which also  $T_{\rm 0}$ is obtained. 
A simplified solution of $T_{\rm 0}$ is given by Eq.~(3) from which an isotope effect on $T_{\rm 0}$ is derived with the parameter values of 
Ref.~3 for $g_{\rm 2}$, $g_{\rm 4}$, and $f$ as appropriate for the $x$ = 1/8 doping, yielding $^{16}$$T_{\rm 0}$ ${\simeq}$ 54.8 K and $^{18}$$T_{\rm 0}$ ${\simeq}$ 57.0 K. 
This corresponds to an OIE exponent of $\alpha_{T_{\rm 0}}$ ${\simeq}$ -0.39, 
in good agreement with the experimental value $\alpha_{T_{\rm LTT}}$ = -0.36(5). Importantly, the calculated values for $^{16}$$T_{\rm 0}$ and $^{18}$$T_{\rm 0}$  
are consistent with the experimental values $^{16}$$T_{\rm LTT}$ and $^{18}$$T_{\rm LTT}$ given above.
The Z-point related soft phonon mode is derived within the SPA from the equations of motion for the two isotopes.
Fig.~5a shows the calculated temperature dependence of the Z-point related phonon mode frequency ${\omega}_{\rm Z}$.  
In this case the transition temperatures are $^{16}$$T_{\rm 0}$ ${\simeq}$ 52.1 K and $^{18}$$T_{\rm 0}$ ${\simeq}$ 54.8 K, which
are slightly shifted to lower temperatures as compared to the experimental values.
However, the corresponding OIE exponent $\alpha_{T_{\rm 0}}$ ${\simeq}$ -0.49 is in agreement with the experiment. In order to achieve an additional correspondence between experiment and theory, the relative mean square Cu-O displacement ${\delta}x$ along the $c$-axis is calculated for both isotopes, $^{16}$O and $^{18}$O, respectively,  and 
their relative difference ${\Delta}$($\delta{x}$)/$\delta{x}$ = ($^{18}$$\delta{x}$-$^{16}$$\delta{x}$)/$^{16}$$\delta{x}$  is shown in Fig. 5b. 
The relative displacement for $^{18}$O is smaller than for $^{16}$O in agreement with the data for the lattice parameter $c$. 
In Fig.~5b the temperature dependence of the relative isotope shift ${\Delta}$($\delta{x}$)/$\delta{x}$ is shown together with ${\Delta}$$c$/$c$ taken from Fig.~2b.
While a temperature independent systematic shift in absolute values is visible, there is salient agreement in the corresponding temperature dependences,
providing strong evidence that the LTT structural phase transition is driven by phonon mode softening 
caused by anharmonic electron-lattice interactions \cite{Annette,Mahan,Zhong,Annette2,Hirsch}.

  In conclusion, oxygen isotope effects on lattice properties in the static stripe phase of LBCO-1/8  were investigated by means of neutron diffraction. The low-temperature tetragonal phase transition temperature $T_{\rm LTT}$
exhibits a large negative OIE with $\alpha_{T_{\rm LTT}}$ = - 0.36(5). 
In addition, the temperature dependent contraction of the $c$-axis with the heavier isotope is observed. 
Since the LTT transition is linked to static charge order, we expect an OIE to be present for the CO temperature as well. 
These results combined with the previously observed OIE on $T_{\rm so}$ evidence that  all transitions observed in the stripe phase of LBCO-1/8 are sensitive to oxygen lattice vibrations. The theoretical lattice dynamical calculations based on 
a dynamic anharmonic electron lattice interaction Hamiltonian consistently reproduce the experimental results. This structural instability is
driven by phonon mode softening stemming from anharmonic electron-lattice interactions, which  
are essential for the stripe formation in the cuprates. 
%The present results may contribute to a better understanding of the complex microscopic mechanism of stripe formation and of high-temperature superconductivity in the cuprates in %general.

%This indicates that the electron-lattice interaction plays an essential role for the stripe formation in cuprate HTS's. 
%Furthermore, the observed oxygen-isotope shifts of the superconducting transition temperature $T_{\rm c1}$ and the
%spin-ordering temperature $T_{\rm so}$ have almost the same magnitude, but opposite signs.
%This provides clear evidence that bulk superconductivity and static spin-order
%are competitive phenomena in the stripe phase of LBCO-1/8, and that the electron-lattice interaction is a  
%crucial factor controlling this competition. The present results may contribute to a better
%understanding of the complex microscopic mechanism of stripe formation and of high-temperature superconductivity in the %cuprates
%in general.

We acknowledge Elvezio Morenzoni, Vladimir Pomjakushin and Jonathan White for helpful discussions. 
%The neutron powder diffraction experiments were carried out at the Swiss spallation neutron source SINQ at PSI, %Switzerland and the ${\mu}$SR experiments were performed at the Swiss Muon Source (S${\mu}$S) at PSI.
This work was supported by the Swiss National Science Foundation, the NCCR MaNEP, 
the SCOPES grant No. IZ74Z0-137322, and the Georgian National Science Foundation grant RNSF/AR/10-16.

\newpage
%%%%%%%%%%%%%%%%%%%%%%%%%%
%\documentclass[aps,prb,twocolumn,showpacs,preprintnumbers,amsmath,amssymb,superscriptaddress]{revtex4}%
%\documentclass[aps,prb,preprint,showpacs,preprintnumbers,amsmath,amssymb,superscriptaddress]{revtex4}%
%\documentclass[aps,prl,preprint,showpacs,preprintnumbers,amsmath,amssymb,superscriptaddress,folatfix]{revtex4}%
%\documentclass[aps,prl,twocolumn,showpacs,preprintnumbers,amsmath,amssymb,superscriptaddress]{revtex4}%
%\documentclass[twocolumn,aps,prl,amssymb,amsfonts,amsmath,color,showpacs]{revtex4}
%\documentclass[aps,prb,preprint,amssymb,showpacs,superscriptaddress]{revtex4}
%\documentclass[twocolumn,showpacs,preprintnumbers,amsmath,amssymb]{revtex4}
%\documentclass[preprint,showpacs,preprintnumbers,amsmath,amssymb]{revtex4}
% Some other (several out of many) possibilities
%\documentclass[preprint,aps]{revtex4}
%\documentclass[twocolumn,showpacs]{revtex4}
%\documentclass[preprint,aps,draft]{revtex4}
%\documentclass[prb]{revtex4}% Physical Review B

%\usepackage{graphicx}% Include figure files
%\usepackage{dcolumn}% Align table columns on decimal point
%\usepackage{bm}% bold math
%\usepackage{color}
%\linespread{2.3}

%\begin{document}

\preprint{preprint(\today)}
\section{SUPPLEMENTAL MATERIAL}
\title{Oxygen Isotope Effects on Lattice Properties of La$_{2-x}$Ba$_{x}$CuO$_{4}$ ($x$ = 1/8)}

\maketitle

New polycrystalline samples of oxygen-substituted ($^{16}$O/$^{18}$O) La$_{2-x}$Ba$_{x}$CuO$_{4}$ with $x$ = 1/8 were prepared in the same way as described in \cite{GuguchiaPRLv2}. The oxygen isotope enrichment of the samples was determined in situ during isotope exchange using mass spectrometry. The
$^{18}$O enriched samples contain ${\simeq}$ 82(5) ${\%}$ $^{18}$O and ${\simeq}$ 18(5) ${\%}$ $^{16}$O.
Magnetization experiments were performed with  a SQUID magnetometer ($Quantum$ $Design$ MPMS-XL) in a field of 0.5 mT and at temperatures between 2 K and 50 K. 
The ${\mu}$SR experiments were carried out at the ${\pi}$M3 beam line at the Paul Scherrer Institut (PSI) in Switzerland using 
the general purpose instrument (GPS) with a standard veto setup providing a low-background ${\mu}$SR signal.
The ${\mu}$SR time spectra were analyzed using the free software package MUSRFIT \cite{Suter}.

%%%%%%%%%%%%%%%%%%%%%%%%%%%%%%%%%%%%%%%%%%%%%%%%%%%%%%%%%%%%%%%
\begin{figure}[b!]
\centering
\includegraphics[width=1.0\linewidth]{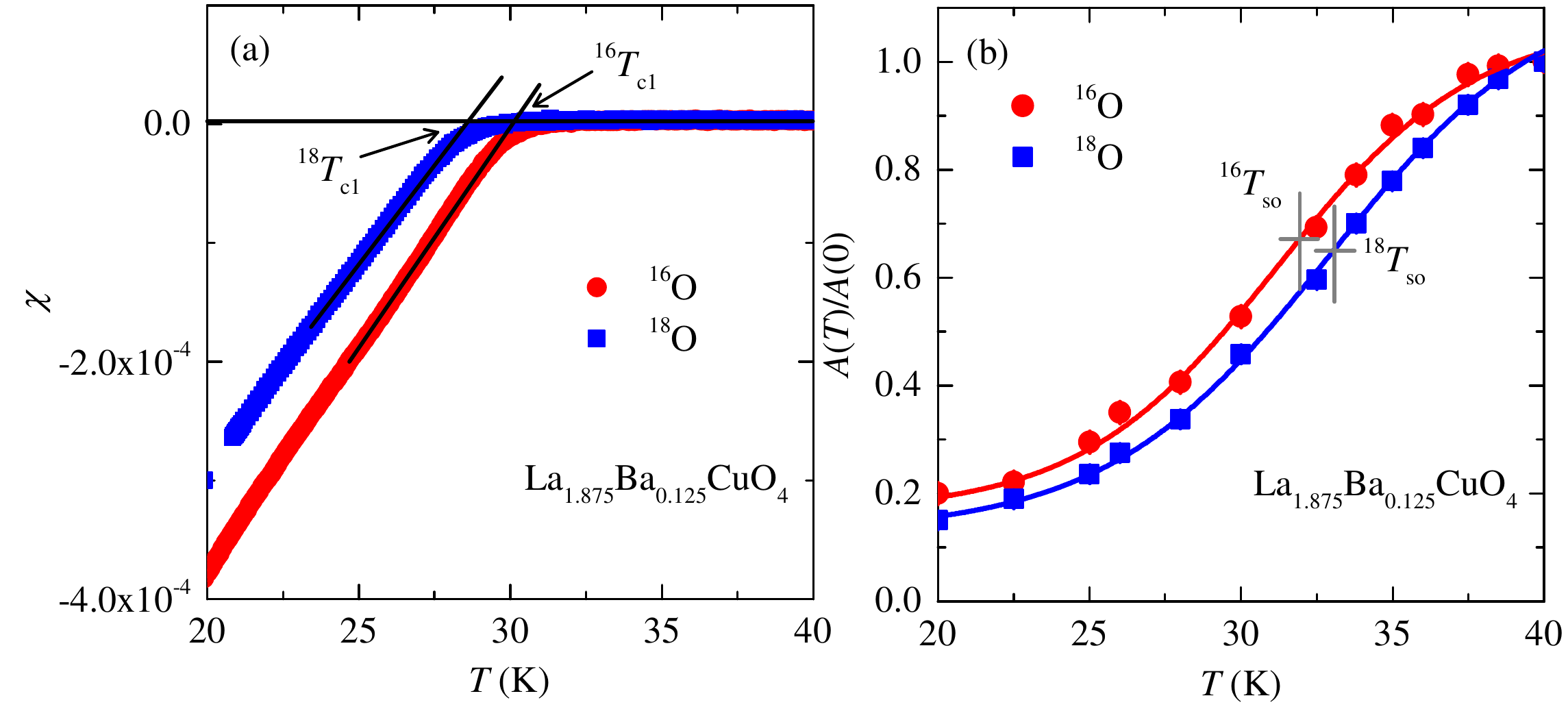}
\vspace{-0.6cm}
\caption{ (Color online) (a) Temperature dependence of the diamagnetic susceptibility $\chi_{\rm ZFC}$
for the $^{16}$O and $^{18}$O samples of LBCO-1/8. The arrows denote the superconducting transition 
temperatures $T_{\rm c1}$. (b)  The normalized transverse-field ${\mu}$SR asymmetry $A$/$A_{\rm 0}$ plotted as 
a function of temperature for the $^{16}$O and $^{18}$O samples  of LBCO-1/8. The crosses 
mark the spin-stripe order temperatures $^{16}$$T_{\rm so}$ and $^{18}$$T_{\rm so}$ 
for the $^{16}$O and $^{18}$O sample, respectively.}
\label{fig1}
\end{figure}
%%%%%%%%%%%%%%%% 

%%%%%%%%%%%%%%%%%%%%%%%%%%%%%%%%%%%%%%%%%%%%%%%%%%%%%%%%%%%%%%%
\begin{figure}[t!]
\centering
\includegraphics[width=1.0\linewidth]{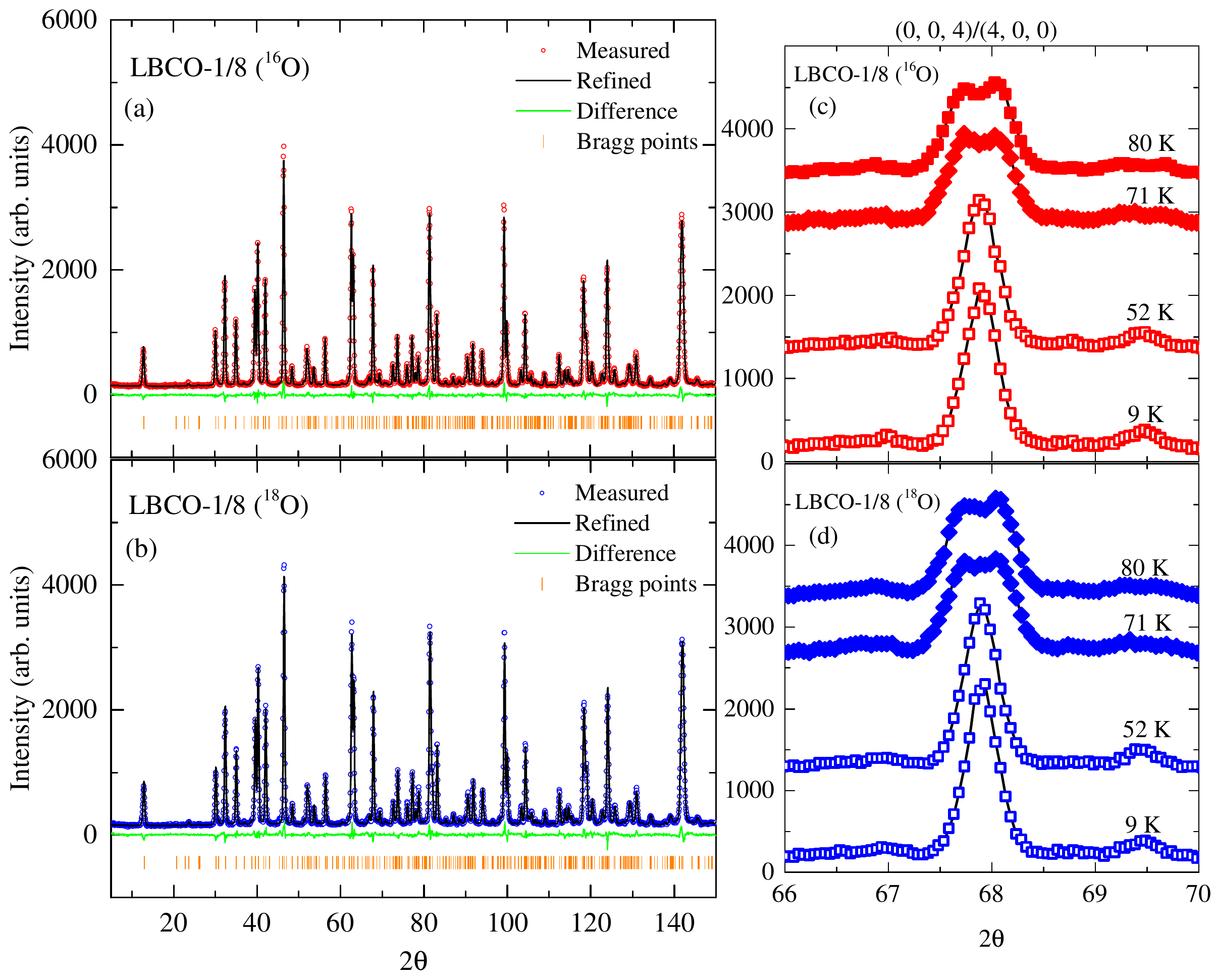}
\vspace{-0.6cm}
\caption{ (Color online) Rietveld refinements of the neutron diffraction patterns for  the $^{16}$O (a) and $^{18}$O (b) samples of LBCO-1/8 at 2 K.
The observed, calculated, and difference plots are shown by open circles, black solid lines, and green solid lines, respectively.
Tick marks below each panel represent the positions of allowed Bragg reflections compatible with space group
$P$4$_{2}$/$ncm$. Panels (c) and (d) show the (400) and (004) Bragg reflections at various temperatures for  the $^{16}$O and $^{18}$O samples, respectively, illustrating the suppression of the orthorhombic splitting below ${\simeq}$ 55 K.}
\label{fig1}
\end{figure}
%%%%%%%%%%%%%%%%% 
The temperature dependence of the zero field cooled (ZFC) diamagnetic susceptibility ${\chi}_{\rm ZFC}$ of the $^{16}$O and $^{18}$O samples  of LBCO-1/8 reveals an analogous behavior as observed in our previous work \cite{Guguchiav2,GuguchiaPRLv2}, namely a two-step superconducting transition \cite{Tranquada2008v2} where at $T_{\rm c1}$ ${\simeq}$ 30 K (the linearly extrapolated susceptibilities intersect the zero line) 2D superconductivity sets in in the CuO$_{2}$ planes, and at $T_{\rm c2}$ ${\simeq}$ 5 K a transition to 3D superconductivity with enhanced diamagnetic response takes place. As discussed previously it is possible to quantify the OIE only for the so called high temperature SC transition $T_{\rm c1}$. Accordingly we show in Fig.~6a  ${\chi}_{\rm ZFC}(T)$ close to $T_{\rm c1}$. Upon replacing $^{16}$O by $^{18}$O, $T_{\rm c1}$ decreases by ${\simeq}$ 1.4 K, corresponding to an OIE exponent $\alpha_{T_{\rm c1}}$ = 0.46(6) in perfect agreement with our previous results \cite{GuguchiaPRLv2}. The transverse field ${\mu}$SR asymmetry $A$ (normalized to its maximum value $A_{\rm 0}$) has been measured in an applied field of ${\mu}_{0}$$H$ = 3 mT following the procedure given in Ref.~\cite{GuguchiaPRLv2} (see Fig.~6b). Above 40 K, $A$ saturates consistent with the fact that the whole sample is paramagnetic. Below 40 K, $A$ systematically decreases signaling the appearance of magnetic order. As already reported previously, $A(T)$ is shifted to higher temperatures for the $^{18}$O sample as compared to the $^{16}$O sample, with an OIE exponent $\alpha_{T_{\rm so}}$ = -0.56(9) in agreement with Ref. \cite{GuguchiaPRLv2}. 

During the Neutron diffraction experiments, the $^{16}$O and $^{18}$O samples of LBCO-1/8 were each placed 
in 8-mm-diameter vanadium containers and measured in the temperature range from 2 K to 300 K. 
The experiments were done using a sample changer inside the cryostat to ensure identical temperature 
conditions for the $^{16}$O  and $^{18}$O samples. 
Neutron powder diffraction patterns recorded at $T$ = 2 K and the corresponding structural refinements
for the $^{16}$O  and $^{18}$O samples of LBCO-1/8 are shown in Fig. 7a and b. 
The average structure was assessed through Rietveld refinements to the raw diffraction data using the program Fullprof \cite{Rodrigezv2} employing  $I$4/$mmm$ (HTT), $Bmab$ (LTO), and $P$4$_{2}$/$ncm$ (LTT) models from the literature \cite{HuckerPRBv2}. 
Figure 7c and d shows that the orthorhombic splitting between the (400) and (004) Bragg reflections is clearly resolved. Below ${\simeq}$ 55 K the orthorhombic splitting is suppressed, signaling the transition to the LTT phase.

%We are grateful to A.~Bussmann-Holder for valuable discussions. 
%The ${\mu}$SR experiments were performed at the Swiss Muon Source, Paul Scherrer Institute (PSI),
%Villigen, Switzerland. 
%This work was supported by the Swiss National Science Foundation, the NCCR MaNEP, 
%the SCOPES grant No. IZ74Z0-137322, and the Georgian National Science Foundation grant RNSF/AR/10-16.


\begin{thebibliography}{22}
\bibitem{Comment} Note that some peculiar features close to $T_{\rm LTT}$ are visible. However, due to the
fact that close to the LTT to LTO structural phase transition both phases coexist (see discussion below) these pecularities are likely related to artefacts 
caused by the exchange of the data treatment procedure.
\end{thebibliography}
\end{document}